\documentclass[9pt,sigplan]{acmart}

\setcopyright{none} 
\renewcommand\footnotetextcopyrightpermission[1]{}

\acmConference[DAC 2026]
{The 62nd ACM/IEEE Design Automation Conference}
{July 2026}
{Long Beach, CA, USA}

\usepackage{fancyhdr}
\usepackage[normalem]{ulem}
\usepackage{url}
\usepackage{tikz}
\usepackage{amsthm}
\usepackage{amsmath,amsfonts}
\usepackage{graphicx}
\usepackage{textcomp}
\usepackage[table,xcdraw]{}
\usepackage{booktabs} % For formal tables
\usepackage{stfloats}
\usepackage{wrapfig}
\usepackage{multirow}
\usepackage{paralist}
\usepackage{balance}
\usepackage{bm}
\usepackage{pifont}
\usepackage{ragged2e}
\usepackage{subcaption}
\usepackage{url}
\usepackage{xspace}
\usepackage{algorithmic}
\usepackage[ruled,linesnumbered]{algorithm2e}
\usepackage{diagbox}
\usepackage{threeparttable}
\usepackage{makecell}
\usepackage{colortbl}
\usepackage{bm}
\definecolor{grey}{RGB}{0.5,0.5,0.5}
\usepackage{tcolorbox}
\usepackage[dvipsnames]{xcolor}
\usepackage{enumitem}
\usepackage{listings} 
\newcommand{\parlabel}[1]{{\noindent\bf #1}}

\newcommand\name{UVMarvel}

\newrobustcmd*\circled[1]{\tikz[baseline=(char.base)]{
            \node[shape=circle,draw,inner sep=1pt,fill,text=white,minimum size=1em] (char) {\textsf{\small #1}};}}

\usepackage{amssymb}
\usepackage{caption}
\usepackage{colortbl}
\usepackage{tabularx}
\usepackage{textcomp}

\usepackage{ragged2e}

\definecolor{codegreen}{rgb}{0,0.6,0}
\definecolor{codegray}{rgb}{0.5,0.5,0.5}
\definecolor{codepurple}{rgb}{0.58,0,0.82}
\definecolor{backcolour}{rgb}{0.95,0.95,0.92}

\lstdefinelanguage[RISC-V]{Assembler}
{
  alsoletter={.}, % allow dots in keywords
  alsodigit={0x}, % hex numbers are numbers too!
  morekeywords=[1]{ % instructions
    lb, lh, lw, lbu, lhu,
    sb, sh, sw,
    sll, slli, srl, srli, sra, srai,
    add, addi, sub, lui, auipc,
    xor, xori, or, ori, and, andi,
    slt, slti, sltu, sltiu,
    beq, bne, blt, bge, bltu, bgeu,
    j, jr, jal, jalr, ret,
    scall, break, nop, li, bset, bext, bins
  },
  morekeywords=[2]{ % sections of our code and other directives
    .align, .ascii, .asciiz, .byte, .data, .double, .extern,
    .float, .globl, .half, .kdata, .ktext, .set, .space, .text, .word
  },
  morekeywords=[3]{ % registers
    zero, ra, sp, gp, tp, s0, fp,
    t0, t1, t2, t3, t4, t5, t6,
    s1, s2, s3, s4, s5, s6, s7, s8, s9, s10, s11,
    a0, a1, a2, a3, a4, a5, a6, a7,
    ft0, ft1, ft2, ft3, ft4, ft5, ft6, ft7,
    fs0, fs1, fs2, fs3, fs4, fs5, fs6, fs7, fs8, fs9, fs10, fs11,
    fa0, fa1, fa2, fa3, fa4, fa5, fa6, fa7
  },
  morecomment=[l]{;},   % mark ; as line comment start
  morecomment=[l]{\#},  % as well as # (even though it is unconventional)
  morestring=[b]",      % mark " as string start/end
  morestring=[b]'       % also mark ' as string start/end
}

\definecolor{mauve}{rgb}{0.58,0,0.82}

\lstdefinestyle{riscv}{
  literate={ö}{{\"o}}1
           {ä}{{\"a}}1
           {ü}{{\"u}}1,
  basicstyle=\footnotesize\ttfamily,                 % very small code
  breaklines=true,                           % break long lines
  commentstyle=\itshape\color{green!50!black}, % comments are green
  keywordstyle=[1]\color{blue!80!black},    % instructions are blue
  keywordstyle=[2]\color{orange!80!black},  % sections/other directives are orange
  keywordstyle=[3]\color{red!50!black},      % registers are red
  stringstyle=\color{mauve},        % strings are from the telekom
  identifierstyle=\color{teal},  % user declared addresses are teal
  language=[RISC-V]Assembler,           % all code is RISC-V
  tabsize=2,                           % indent tabs with 2 spaces
  showstringspaces=false  % do not replace spaces with weird underlines
  breakatwhitespace=false,         
  breaklines=true,                 
  captionpos=b,                    
  keepspaces=true,                 
  numbers=left,                    
  numbersep=5pt, 
  xleftmargin=10pt,
  frame=single,
  showspaces=false,                
  showtabs=false,                  
  numberblanklines=false
}
   \usepackage{tikz}
   \usetikzlibrary{positioning}
\begin{document}

\definecolor{darkgreen}{rgb}{0, 0.6, 0}
% {RGB}{66, 186, 71}

\newcommand{\approach}{UVmarvel}
% \newcommand{\name}{LiFU}

% \title{
% UVMarvel: an Automated LLM-aided UVM Machine for Subsystem-level RTL Verification%
% \thanks{This paper has been accepted by DAC 2026 and will appear in the proceedings.}
% }
\title{UVmarvel: an Automated LLM-aided UVM Machine for
Subsystem-level RTL Verification}

\titlenote{This paper has been accepted by DAC 2026 and will appear in the proceedings.}

\author{
Junhao Ye$^{1,2}$, Dingrong Pan$^{2}$, Hanyuan Liu$^{1,2}$, Yuchen Hu$^{1,2}$, Jie Zhou$^{1,2}$, Ke Xu$^{1,2}$,\\ Xinwei Fang$^3$, Xi Wang$^{1,2}$, Nan Guan$^4$, Zhe Jiang$^{1,2}{^\dagger}$
}
% \thanks{$^\dagger$Corresponding author:

% Zhe Jiang. Email: 
% zhejiang.uk@gmail.com.

% Xinwei Fang. Email:
% xinwei.fang@york.ac.uk.

% Xi Wang. Email:
% xi.wang.sudo@gmail.com}
\affiliation{%
  % \institution{        
  % $^1$School of Integrated Circuits, Southeast University, China $^2$National Center of Technology Innovation for EDA, China \\
  %       $^3$National University of Defense Technology, China
  %       $^4$Department of Computer Science, City University of Hong Kong, Hong Kong\\
  %       $^5$Department of Computer Science, University of York, UK\\}
  \institution{        
  $^1$Southeast University, China $^2$National Center of Technology Innovation for EDA, China \\
        $^3$University of York, UK
        $^4$City University of Hong Kong, Hong Kong\\
        $^\dagger$Corresponding author: zhejiang.uk@gmail.com
        }
  \country{}
}

\begin{abstract}
% Verification presents a major bottleneck in Integrated Circuit (IC) development, consuming nearly 70\% of the total development effort. While the Universal Verification Methodology (UVM) is widely used in industry to improve verification efficiency through structured and reusable testbenches, constructing these testbenches and generating sufficient stimuli remain challenging.
% These challenges arise from the considerable manual coding effort required, repetitive manual execution of multiple EDA tools, and the need for in-depth domain expertise to navigate complex designs. Here, we present \textbf{\name}, an automated verification framework that leverages Large Language Models (LLMs) to generate UVM testbenches for subsystem-level RTL and iteratively refine them using coverage feedback, significantly reducing manual effort while maintaining rigorous verification standards.
% \textbf{\name} is the first framework capable of automatically constructing subsystem-level UVM testbenches across mainstream protocols. It achieves an average code coverage of 95.65\% approaching industrial verification standards. 
Verification presents a major bottleneck in Integrated Circuit (IC) development, consuming nearly 70\% of total effort. While the Universal Verification Methodology (UVM) improves reuse through structured verification environments, constructing subsystem-level UVM testbenches and generating high-quality stimuli still require extensive manual coding, repeated EDA tool runs, and deep protocol and micro-architectural expertise. We present \textbf{\name}, an automated verification framework that leverages Large Language Models (LLMs) to build UVM testbenches for subsystem-level RTL. \textbf{\name}\ introduces an Intermediate Representation (IR) and a Bus Protocol Library to translate heterogeneous specifications into protocol-correct subsystem-level UVM testbenches, and employs a Signal Tracker and a Verilog Patching Library to guide LLM-based stimuli refinement. \textbf{\name}\ is the first framework capable of automatically constructing subsystem-level UVM testbenches across mainstream bus protocols, and it achieves an average code coverage of 95.65\%, while reducing verification time from several human working days to a 4.5-hour automated execution.

% approaching industrial verification standards while substantially reducing manual effort.
% and delivering a 20.47× efficiency gain over experienced engineers.
% \textcolor{red}{to be updated}
\end{abstract}

% % ⭐ 在这里写说明（不会报错！）
% \begin{center}
% \footnotesize{* Corresponding authors: xinwei.fang@york.ac.uk, xi.wang@seu.edu.cn, zhejiang.uk@gmail.com}
% \end{center}

\maketitle

\setlength{\textfloatsep}{9pt}% Remove \textfloatsep
\vspace{-10pt}
\section{Introduction}
\label{sc:Intro}
\vspace{-2pt}

\begin{figure}[t]
    \centering
    % 第一个子图
    \begin{subfigure}[b]{\linewidth}\centering
        \includegraphics[width=1\linewidth]{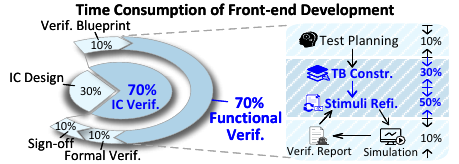}
        \Description{IC Front-end Design.}
        \subcaption{Decomposition of IC front-end design and verification. Functional verification accounts for nearly 70\% of verification effort, starting from test planning, with the most time spent on testbench construction and stimuli refinement.}
        \label{fig:breakdown_ic}
    \end{subfigure}

    % \vspace{-6pt}
    
    % 第二个子图
    \begin{subfigure}[b]{\linewidth}
        \includegraphics[width=1\linewidth]{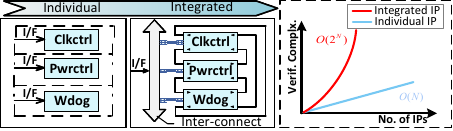}
        \Description{Comparison of verification complexity between individual and integrated IPs.}
        \subcaption{
        % Exponential verification complexity in integrated IP compares to linear scaling for individual IP. The differences between the levels are interface interaction and constraint couplings.
        Comparison of verification complexity between individual and integrated IPs. Interactions and constraint couplings cause growth of complexity during integration.}
        \label{fig:complex_soc}
    \end{subfigure}

    \vspace{-6pt}
    \caption{
    % The IC front-end process comprises design and verification phases}, with verification taking over 70\% of the total effort. Verification complexity of module level grows linearly with the number of modules, since each IP can be verified with its own functions Independently. While at the subsystem level, because of inter-IP interface interaction and constraint couplings, the verification complexity increases exponentially with number of integrated IP modules.
    \textbf{\footnotesize Verification dominates IC front-end development.} While module-level complexity scales linearly, subsystem-level verification grows exponentially due to inter-IP dependencies. (Verif.: Verification, TB: testbench, Refi.: Refinement, Constr.:Construction, I/F: Interface, Complx.: Complexity)}
    \label{fig:workflow}
    \vspace{-2pt}
\end{figure}

% Under the trend of Agile hardware development, modern SoC projects iterate rapidly and face tight delivery schedules. As shown in Fig.\ref{fig:breakdown_ic}, verification has become the dominant bottleneck, accounting for almost 70\% of the total verification cycle \cite{foster2020wilson}. 
% Industrial verification predominantly adopts the Universal Verification Methodology (UVM), a standardized, class-based framework that provides modular agents, layered testbench architecture, and transaction-level abstraction to support scalable and reusable functional verification.
% While UVM provides structure and scalability, constructing a full UVM testbench for real-world designs remains a labour-intensive process that demands significant engineering expertise.
Agile hardware development pushes modern SoC projects toward rapid iteration and frequent design revisions, raising pressure on verification teams to match fast-changing designs. As shown in Fig.\ref{fig:breakdown_ic}, verification has become the dominant bottleneck, consuming nearly 70\% of the overall front-end cycle \cite{foster2020wilson}. 
% To manage this complexity, industrial flows predominantly rely on UVM~\footnote{A standardised, class-based architecture that organises agents, transactions, sequences, and scoreboards into a scalable and reusable verification framework}. Despite its structure, UVM remains manual in practice, and assembling a full testbench for real designs requires deep engineering expertise.
To manage this complexity, industrial flows predominantly rely on UVM. Despite its structure, UVM remains manual in practice, and assembling a full testbench for real designs requires deep engineering expertise.

% Template-driven scripting approaches have long been used to reduce manual verification burden; however, their effectiveness is fundamentally limited. They automate universal tasks—directory setup, template instantiation, compilation rules—but lack the micro-architectural knowledge required to reason about interface roles, timing semantics, hierarchy, and signal dependencies. As a result, large portions of driver/monitor behaviour, functional intent, and design-specific stimulus still require extensive manual work. Since these scripts do not incorporate micro-architectural knowledge, their generated stimuli remain shallow or overly random, providing limited improvement in coverage. 
To reduce this manual burden, companies incorporate template-driven scripting flows on top of UVM\cite{ev2019design, georgoulopoulos2019uvm, pavithran2017uvm}. These automate repetitive and well-structured tasks, e.g., directory initialisation, code skeleton generation, or basic register and sequence templates and thus improve consistency. 
Yet, since these tools expand predefined templates without understanding micro-architectural intent, they cannot infer protocol timing, transaction semantics, driver/monitor behaviour, or functional constraints. As designs evolve, engineers must manually fill these semantic gaps, and coverage closure continues to depend on expert effort rather than automation.

% With the emergence of LLMs, which can process natural-language specifications, RTL code, structural descriptions, verification automation has gained opportunities beyond traditional scripting. Prior works — such as MEIC\cite{xu2024meic}employs dual fine-tuned LLMs with RTL toolchain, to automate syntax/function error detection and correction in Verilog; UVLLM\cite{hu2024uvllm}integrates LLMs with UVM methodology, to automate syntax/function error detection and repair in Verilog code, ensuring comprehensive verification; AssertLLM\cite{yan2025assertllm} utilizes three customized LLMs to generate SystemVerilog Assertions from specification documents for RTL verification. — show that LLMs can assist with debugging and localized test generation. UVM2 represents the first attempt at full-phase UVM automation for IP-level verification, generating agents, environments, sequences, and basic stimuli end-to-end. 

Recent advances in the LLMs create new opportunities for breaking this bottleneck.
Unlike template systems, the LLMs can process natural-language specifications, RTL structure, protocol rules, and design-intent descriptions\cite{li2025specllm, wang2025deepassert, liu2024rtlcoder, fang2402assertllm, maddala2024laag, shih2025flag, bhandari2024llm}, allowing them to reason about behaviours that traditional scripts cannot capture. Prior works, e.g., MEIC\cite{xu2024meic} employs the dual fine-tuned LLMs with RTL toolchain, to automate error detection and correction in Verilog; UVLLM\cite{hu2024uvllm} integrates the LLMs with UVM methodology, to automate error detection and repair in Verilog code, ensuring comprehensive verification; AssertLLM\cite{yan2025assertllm} utilises three customized LLMs to generate SystemVerilog Assertions from specification documents for RTL verification, showing that the LLMs can assist with debugging and localized test generation. 
Among them, UVM²~\cite{ye2025concept} is the first attempt at UVM automation verification, generating agents, sequences, and basic stimuli end-to-end. 
Yet, these advances remain confined to the IP level, where design interactions are local and protocol reasoning is relatively contained.

% \noindent \textbf{Challenges.}
% Extending LLM-aided UVM verification beyond the IP level to subsystem-level RTL creates a new set of challenges across the verification pipeline, with UVM testbench construction and stimulus refinement proving to be the most limiting factors.
% When we utilised the LLM-aided framework to verify an SoC subsystem, we met bottlenecks at almost every stage of the verification pipeline, 

\noindent \textbf{Challenges.} Extending LLM-aided UVM verification beyond the IP level to an SoC subsystem creates a new set of challenges across the verification pipeline, from constructing the UVM testbench to generating stimuli.
Even on a simple subsystem (a P‑channel power controller), the framework repeatedly failed to assemble a UVM testbench once several interface signals were absent (Sec.~\ref{subsc:ex_bus}).
% Fig.~\ref{fig:uvm-fail-example}
    Unlike IP‑level designs, subsystems present richer bus interfaces and more intricate control paths, documented in lengthy specifications with timing waveforms and block diagrams, which cause an exponential growth in complexity with the number of IP modules, as shown in Fig.~\ref{fig:complex_soc}. This multi‑modal documentation currently exceeds the interpretive capacity of the general‑purpose LLMs, making automatic assembly of a correct subsystem-level UVM testbench unreliable.

% Even after we manually constructed the UVM testbench, the generated stimuli still achieved unacceptably low coverage (below 40\%). 

Even when a UVM testbench is available, the generated stimuli still achieve low coverage, average below 40\%. They fail to exercise corner‑case behaviours, long dependency chains across multiple IPs, rare event sequences and intricate state transitions. For example, the subsystem’s behaviour often hinges on multi‑component sequences (e.g., power request/acknowledgement handshakes) that must be orchestrated step by step, whereas the LLMs tend to emit simple, repetitive patterns that quickly saturate obvious cases while missing deeper corners. Without a clear view of inter-IP dependencies, the LLMs cannot reliably derive the nuanced sequences needed to cover all functional points, leading to early convergence of coverage.

\noindent \textbf{Contributions.}
We present \textbf{\name}, the first automated, LLM-driven framework for subsystem-level UVM-based verification. To automatically construct subsystem-level UVM testbenches, \textbf{\name}\ integrates a verification-oriented Intermediate Representation (IR) with a scalable Bus Protocol Library, enabling the LLMs to interpret structural semantics and protocol behaviours across heterogeneous interfaces.
For stimuli refinement, \textbf{\name}\ employs a Signal Tracker that extracts inter-IP dependency paths across modules, and a Verilog Patching Template Library that distils each path into minimal semantic blocks, enabling the LLMs to derive the nuanced multi-step sequences required for coverage improvement.
% As a unified solution that provides subsystem-level structural and behavioural visibility to the LLMs, \textbf{\name}\ achieves industrial-grade coverage of 95.65\%. 
As a unified solution that provides subsystem-level access to both design structure and execution behaviour for the LLMs, \textbf{\name} achieves industrial-grade coverage of 95.65\%.
The Bus Protocol Library and Verilog Patching Template Library are open-sourced at 
https://github.com/SEU-ACAL/reproduce-UVMarvel-DAC-26.

\begin{figure*}[t]
    \includegraphics[trim=01mm 01mm 18mm 02mm, clip, width=0.9\linewidth]{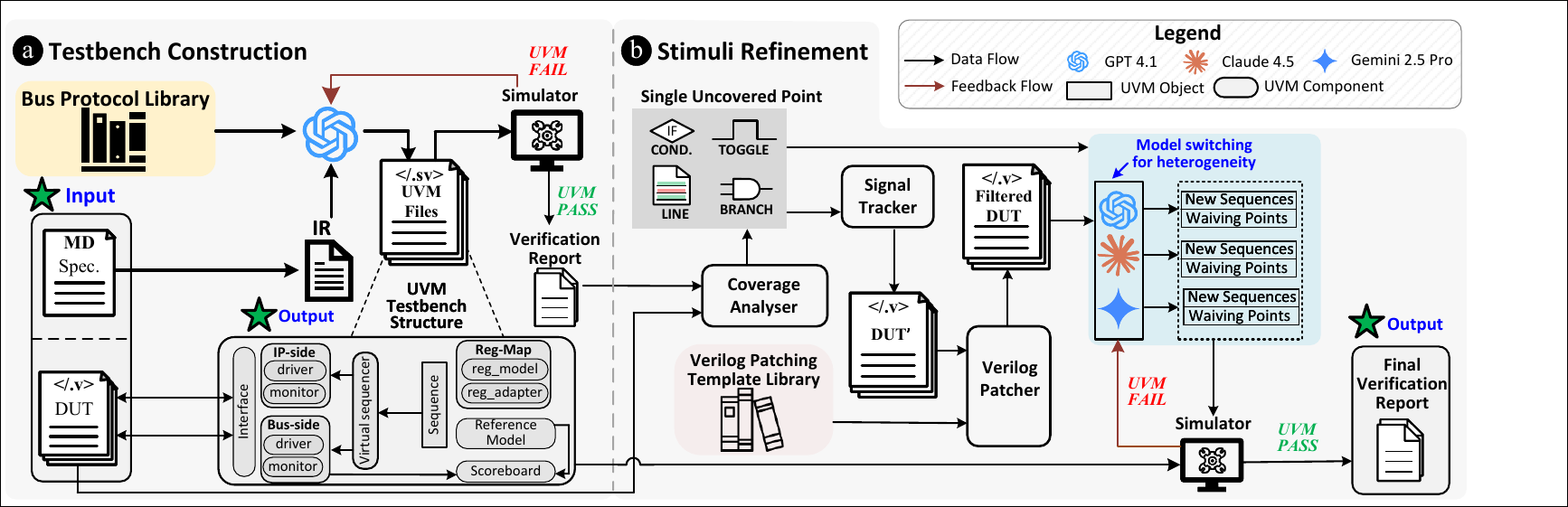} 
    %\descriptionlabel{overview}
    \vspace{-5pt}
    \caption{
    \textbf{\footnotesize {\name} Framework.} (a) \textbf{Testbench Construction}: the IR translated from design specifications, together with the Bus Protocol Library, guiding the LLMs to construct UVM testbench; and (b) \textbf{Stimuli Refinement}: uncovered coverage data are interpreted by the Coverage Analyser, filtered DUT is identified through Signal tracker and Verilog Patcher, and the LLMs generate new stimuli or waiving points to improve coverage.
    }
    \vspace{-10pt}
    \label{fig:stage} % 便于后续引用
\end{figure*}

\vspace{-5pt}
\section{\name: An Overview}
\label{sc:overview}
\vspace{-2pt}

Aiming to accelerate the verification loop for subsystem-level RTL, \name\ integrates LLM assistance into a standard UVM workflow and organises the process into two stages, as shown in Fig.~\ref{fig:stage}. The framework takes specifications and RTL as inputs and constructs a complete UVM testbench with final verification reports that include error logs, waiving candidates, and coverage results.

Stage \circled{a} aims to construct testbenches. Information about the subsystem is often dispersed across specifications and RTL, making it difficult for the LLMs to understand the structure and interface behaviours. To provide a unified view, we convert these materials into an IR that summarises modules, connections and interface roles. Since the subsystem's behaviour further depends on protocol rules that the LLMs cannot easily extract, a Bus Protocol Library is supplied to give precise transaction guidance. With these supports, the LLMs generate the UVM testbenches, executed with the DUT to produce the first verification report.

Stage \circled{b} aims to improve coverage. The problem is that initial stimuli typically fail to exercise deep, multi-IP interactions and long dependency chains, causing early coverage convergence. To address this, the coverage analyser decomposes the report into single uncovered points. Each point is examined by the Signal Tracker, which identifies the relevant signal path in the RTL. A patching step then patches the Verilog structure to produce a compact slice that preserves the essential dependency chain to generate a filtered DUT (a DUT that has only key path-related code). This filtered DUT, combined with the uncovered point, provides the context needed for the LLMs to generate additional stimuli or justified waiving candidates. The new stimuli are added, and the testbench is re-executed on the original DUT to improve coverage, while also producing the final verification report.

% Both stages rely on the LLMs to synthesise UVM components and stimuli, and any artefacts that fail to compile trigger a feedback mechanism for regeneration, ensuring robust automation.

%\input{contents/motivate}
% \input{contents/Approach}
\vspace{-5pt}
\section{\name: The Framework Pipeline}
\label{sc:pipeline}

\subsection{Intermediate Representation}

Verification in practice typically begins with a test plan (Fig.~\ref{fig:breakdown_ic}), where engineers interpret the specification and decompose the design intent into concrete verification points. Existing LLM-based approaches, yet, devote little effort to this planning stage~\cite{chen2024llm4dv,xu2024specllm,berabi2024llm4hw}. 

% In real designs, subsystem specifications mix block diagrams and timing waveforms (Fig.~\ref{fig:ir}) and are written to describe functional behaviour rather than to drive verification, so extra work is needed to decide which UVM components should be built from the specification.

From the perspective of a UVM testbench, what we need is information that is much more concrete than what the specification exposes: (i) for the environment hierarchy (\texttt{uvm\_env}, \texttt{uvm\_agent}, \texttt{uvm\_driver}, \texttt{uvm\_monitor}), it must know how the DUT is instantiated and how it communicates with the outside system; (ii) for the \texttt{uvm\_reg\_block} and its adapter, it needs the programmable state (which registers exist, where they are mapped, how they are reset and accessed); and (iii) for sequence generation, it needs timing assumptions on the interfaces and the key functional scenarios and corner cases that should be exercised. In practice, human verification engineers gather exactly this information before they start writing UVM code or a test plan.

\begin{figure}[t]    
\centering    
\vspace{5pt}
\includegraphics[width=1\linewidth]{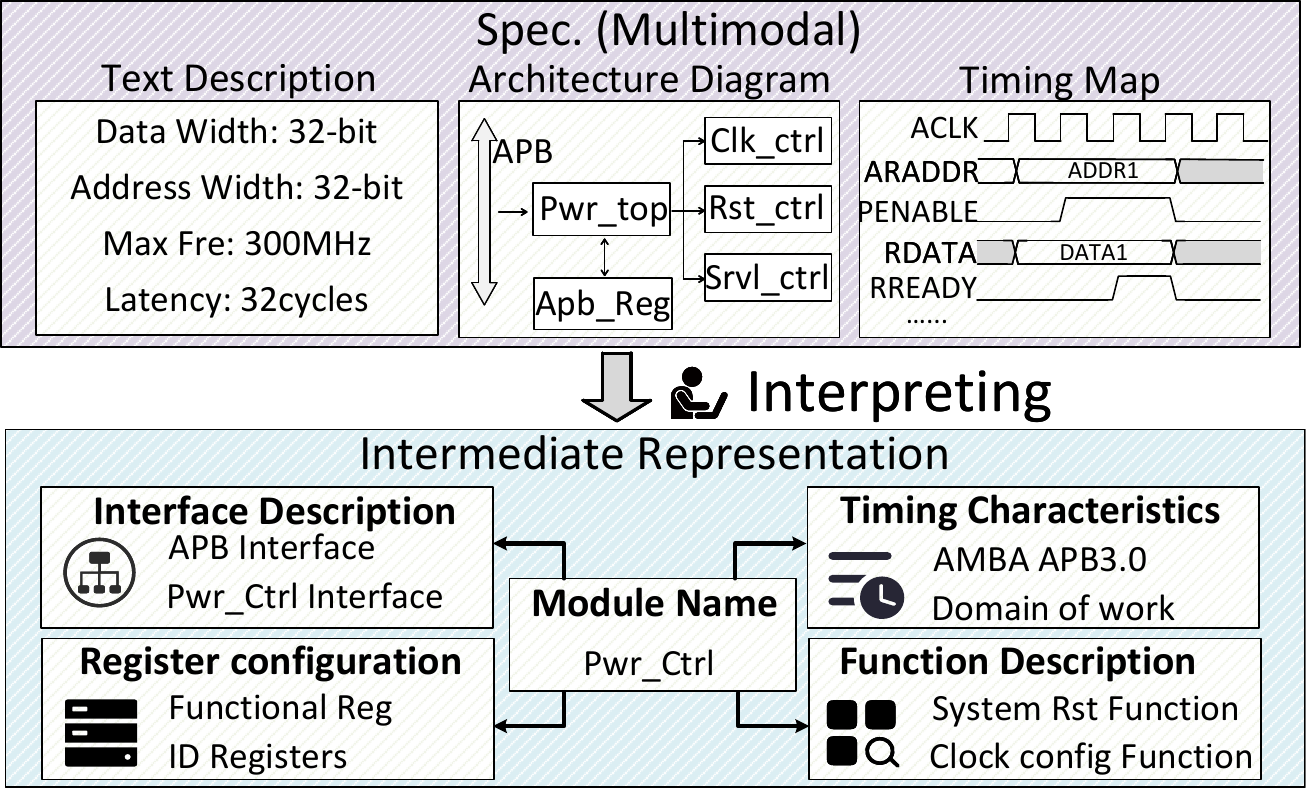}    
\Description{IR and spec.}    
\vspace{-15pt}    
\caption{
\textbf{\footnotesize Multimodal specification vs. unified IR.} IR condenses diagrams, waveforms and textual descriptions into verification-relevant fields.
}    
\vspace{-5pt}    
\label{fig:ir}
\end{figure}

Guided by this observation and standard verification methodology documents~\cite{uvm12manual,spear2012systemverilog,bergeron2000writing,cadenceMDV}, we construct a five-part IR with components \emph{Module Name}, \emph{Interface Description}, \emph{Register Configuration}, \emph{Timing Characteristics} and \emph{Functional Description}, as shown in Fig.~\ref{fig:ir}. This IR distils the information needed for UVM environment, agent, register model and scenario construction into a verification-centred description that the LLMs can use directly.

\vspace{-10pt}
\subsection{Bus Protocol Library}
\label{subsc:bus_lib}
When constructing a subsystem-level UVM testbench, bus-side components, e.g., drivers, monitors, agents and interfaces are indispensable: without them, the subsystem cannot interconnect its IP blocks or exercise its behaviour end-to-end. Industrial verification IP (VIP) packages protocol rules into reusable driver, monitor and checker~\cite{melikyan2021uvm,vagaggini2022spacewire,harutyunyan2020configurable}, but these VIPs neither synthesise new, design-specific bus components from heterogeneous specifications nor expose protocol structure in a form that the LLMs can adapt. \footnote{In our preliminary SoC experiments(Sec.\ref{subsc:ex_bus}), a general-purpose LLM asked to construct bus-side components directly repeatedly failed to assemble a correct subsystem-level UVM testbench once some interface signals were omitted.}

Mainstream bus protocols encode rich ordering rules, handshake relations and timing dependencies, and their official specifications are long and heterogeneous, often mixing timing diagrams, waveforms and descriptive text. Even with these documents available, current LLMs struggle to turn protocol manuals into correct UVM components, due to limited context, long-range dependencies and multi-modal artefacts~\cite{chen2024multi}. At the same time, these protocols rely on a small set of regular behavioural patterns, and the LLMs are effective at reading and modifying structured code when guided by constraints. This motivates representing protocol behaviours as a concise set of human-readable UVM skeletons that capture the stable request/response and timing rules, and then asking the LLMs to specialise these skeletons to each DUT.

\begin{figure}[t]
    \centering
    \vspace{-5pt}
    \includegraphics[trim=7mm 7mm 7mm 0mm, clip, width=0.75\linewidth]{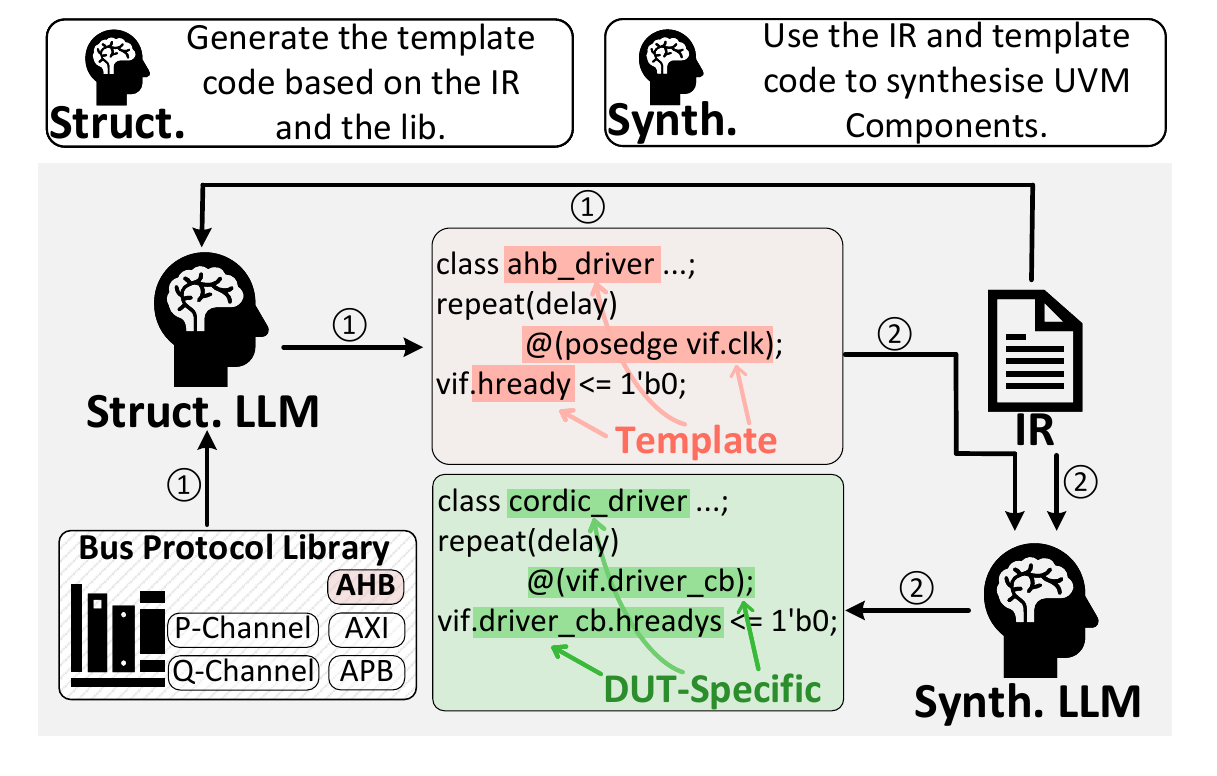} 
    \Description{llm generation flow}
    \vspace{-10pt}
    \caption{
    \textbf{\footnotesize Generation of UVM bus components using the Bus Protocol Library, illustrated with AHB\_Driver.} It selects a protocol-specific UVM skeleton, then the LLM specialises it into DUT-specific code that is integrated as a key bus component.(Struct.: Structural, Synth.: Synthetic)
    }
    \vspace{-10pt}
    \label{fig:bus_lib}
\end{figure}

Based on this idea, we built a Bus Protocol Library to assist our framework in generating bus-side UVM components. For each supported protocol, the library provides UVM skeletons for the interface, driver, monitor and agent. During generation, the framework uses the protocol information in the IR to select the skeletons and prompts the LLMs to specialise them with the remaining IR details, such as signal names, groupings, widths and address ranges, as illustrated in Fig.~\ref{fig:bus_lib}. To keep protocol semantics consistent across designs, the LLMs are allowed to modify only designated regions of the skeletons (for example, configuration parameters), while the protocol control flow and handshake structure remain fixed. In this way, the protocol semantics come from the library, and the LLMs mainly adapt them to each DUT instance.

\vspace{-5pt}
\subsection{Coverage Analyser}
\label{subsc:cov_analy}

After the UVM testbench has been constructed with the assistance of IR and the Bus Protocol Library, the verification flow proceeds to simulation. Each run produces a coverage report, typically as large HTML pages that mix coverage data with UI markup. This format is convenient for human browsing but unsuitable for the LLMs, since most tokens are layout noise and the useful coverage items are scattered across many sections. To make coverage feedback usable, we introduce a Coverage Analyser that converts these reports into a compact, task-oriented summary.

The analyser takes the HTML coverage report as input and parses the coverage categories. For each category, it extracts all uncovered or partially covered items plus essential context such as hierarchical name and source file. It then emits these items as structured text grouped by coverage type and module, and feeds them to the LLMs. This strips away UI boilerplate and surfaces only uncovered targets, so the LLMs can focus on generating additional stimuli to close the remaining gaps.

{\renewcommand{\baselinestretch}{0.6}
\selectfont
\small
\begin{algorithm}[b]
\caption{Single-File Signal Tracing}
\label{al:single}
\KwIn{Target signals $S_0$, Verilog file $F$}
\KwOut{Relevant statement set $\mathcal{R}$}
$Q \leftarrow S_0$, $V \leftarrow \emptyset$, $\mathcal{R} \leftarrow \emptyset$\;
\While{$Q \neq \emptyset$}{
  $s \leftarrow \mathrm{dequeue}(Q)$\;
  \lIf{$s \in V$}{continue}
  $V \leftarrow V \cup \{s\}$\;
  \ForEach{statement $t$ in $F$ referencing $s$}{
    $\mathcal{R} \leftarrow \mathcal{R} \cup \{t\}$\;
    $X \leftarrow \text{signals in } t \setminus \{s\}$\;
    \lForEach{$x \in X \setminus V$}{enqueue($Q,x$)}
  }
}
\Return{$\mathcal{R}$}
\end{algorithm}
}

\vspace{-1.5pt}
\subsection{Signal Tracker} 
\label{subsc:ST}

After coverage feedback has been collected, UVMarvel analyzes why certain coverage points remain uncovered. In practice, the LLM-generated stimuli often leave many deep corner cases untested, resulting in low coverage.
The missing cases are usually deep corners whose activation depends on long signals chains across several IPs and many cycles, such as handshakes. Hitting these points requires driving the right signals in a specific order and times. In subsystem-level RTL, the relevant assignments and conditions are scattered over many files and always blocks, so the LLMs cannot easily see how an uncovered signal is connected back to controllable inputs.
% After we have a working subsystem-level UVM testbench, the LLM-generated stimuli often remain at low coverage. 
% The missing cases are usually deep corners whose activation depends on long chains of signals across several IPs and over many cycles, such as power request sequences. Hitting these points requires driving the right signals in a specific order and at specific times. In subsystem-level RTL, the relevant assignments and conditions are scattered over many files and always blocks, so the LLMs cannot easily see how an uncovered signal is connected back to controllable inputs.

Existing work such as UVLLM~\cite{hu2024uvllm}, inspired by STRIDER~\cite{yang2023strider}, uses AST-based dependency trees to locate important signals, but these trees grow large on subsystem-level RTL and are not centred on any specific coverage point. We instead follow how an engineer debugs coverage: start from the uncovered signal, collect the statements that define or use it, and trace backwards to the top-level I/Os. The Signal Tracker automates this process: it takes as input a set of seed signals $S_0$ from uncovered coverage expressions and the subsystem-level Verilog files, and produces a compact cross-file dependency slice $\mathcal{G}$ together with the top-level I/O ports through which a testbench can drive these signals.

% {\renewcommand{\baselinestretch}{0.6}
% \selectfont
% \small
% \begin{algorithm}[b]
% \caption{Single-File Signal Tracing}
% \label{al:single}
% \KwIn{Target signals $S_0$, Verilog file $F$}
% \KwOut{Relevant statement set $\mathcal{R}$}
% $Q \leftarrow S_0$, $V \leftarrow \emptyset$, $\mathcal{R} \leftarrow \emptyset$\;
% \While{$Q \neq \emptyset$}{
%   $s \leftarrow \mathrm{dequeue}(Q)$\;
%   \lIf{$s \in V$}{continue}
%   $V \leftarrow V \cup \{s\}$\;
%   \ForEach{statement $t$ in $F$ referencing $s$}{
%     $\mathcal{R} \leftarrow \mathcal{R} \cup \{t\}$\;
%     $X \leftarrow \text{signals in } t \setminus \{s\}$\;
%     \lForEach{$x \in X \setminus V$}{enqueue($Q,x$)}
%   }
% }
% \Return{$\mathcal{R}$}
% \end{algorithm}
% }

\textbf{Single-file tracing.}
Algorithm~\ref{al:single} works on a Verilog file $F_j$. Given the current seed set $S$, it finds all statements that read from or assign to any signal in $S$, adds those statements to $\mathcal{R}$, and pushes any newly seen signals into the queue. The resulting set $V_j$ is a small subset of RTL directly related to $S$ within file $F_j$.

Subsystem-level designs usually span many \texttt{.v} files, and important chains often cross module boundaries. We therefore extend the tracing across files using a simple recursive expansion:

{\renewcommand{\baselinestretch}{0.6}
\selectfont
\begin{algorithm}[t]
\small
\caption{Cross-File Recursive Tracing for Subsystems}
\label{al:cross}
\KwIn{\hbox{Target signals $S_0$, submodule files $\{F_1,\dots,F_k\}$, top-level file $F_{\text{top}}$}} 
\KwOut{Global dependency set $\mathcal{G}$}
$S \leftarrow S_0$, $i \leftarrow 0$, $\mathcal{G} \leftarrow \emptyset$\;
\Repeat{$S = \emptyset$}{
  $i \leftarrow i+1$, $E \leftarrow \emptyset$\;
  \ForEach{submodule file $F_j$}{
    $V_j \leftarrow \text{Algorithm~\ref{al:single}}(S,F_j)$\;
    $\mathcal{G} \leftarrow \mathcal{G} \cup V_j$\;
    extract I/O signals of $F_j$ from $V_j$ and add to $E$\;
  }
  $S \leftarrow \text{unique}(E)$\;
}
Run Algorithm~\ref{al:single} on $F_{\text{top}}$ using $\mathcal{G}$ and keep all I/O ports of $F_{\text{top}}$\;
\Return{$\mathcal{G}$}
\end{algorithm}
}

\textbf{Cross-file expansion.}
Starting from $S_0$, we apply Algorithm~\ref{al:cross} to each submodule file $F_j$ and merge the fragments $V_j$ into the global set $\mathcal{G}$. From each $V_j$ we keep only the submodule input and output ports, collect them into a new seed set $S$, and repeat until no new interface signals are found. Finally, we run Algorithm~\ref{al:single} on the top-level file $F_{\text{top}}$ using the accumulated dependencies in $\mathcal{G}$ and keep all top-level I/O ports as legal stimulus entry points.

At this stage, the tracker has identified the RTL statements and signals related to each uncovered coverage point and linked them to controllable top-level interfaces. These statements are still fragments, scattered across files and missing their original module and process context. In the next step, we reconstruct a small amount of Verilog structure around them so that they form a coherent view of the design that can be fed to the LLMs.

\vspace{-5pt}
\subsection{Verilog Patcher} 
\label{subsubsc:VPTL}

As described in Section~\ref{subsc:ST}, the Signal Tracker identifies RTL statements and signals causally connected to each uncovered coverage point and links them to the relevant top-level interfaces. However, once lifted from their original files, these statements are scattered across different always blocks and modules and often lose the headers and surrounding control that make them valid Verilog. Although LLMs can tolerate some incompleteness and edit partially structured programs~\cite{yin2023codetransform,li2024large}, they still need basic syntactic units to understand control and data flow.

To restore this minimal structure, we introduce the Verilog Patching Template Library and a Patcher that uses it. The library provides a small set of patterns for canonical RTL constructs, including module shells, always blocks, case blocks, continuous assignments and instance-level connections, each preserving only the essential syntactic form. During patching, statements that originated from the same construct are grouped and attached to the corresponding template. For example, if the tracker collects branch assignments from a case block but the surrounding \texttt{case}/\texttt{endcase} are missing, the Patcher applies a case-block template to rebuild the header and closing keyword; if an assignment was originally inside an \texttt{always} block but is now isolated, the Patcher uses an always-block template to reintroduce the minimal process wrapper.

\begin{figure}[h]
    \vspace{-5pt}
    \includegraphics[trim=3mm 4mm 2mm 1mm, clip, width=0.8\linewidth]{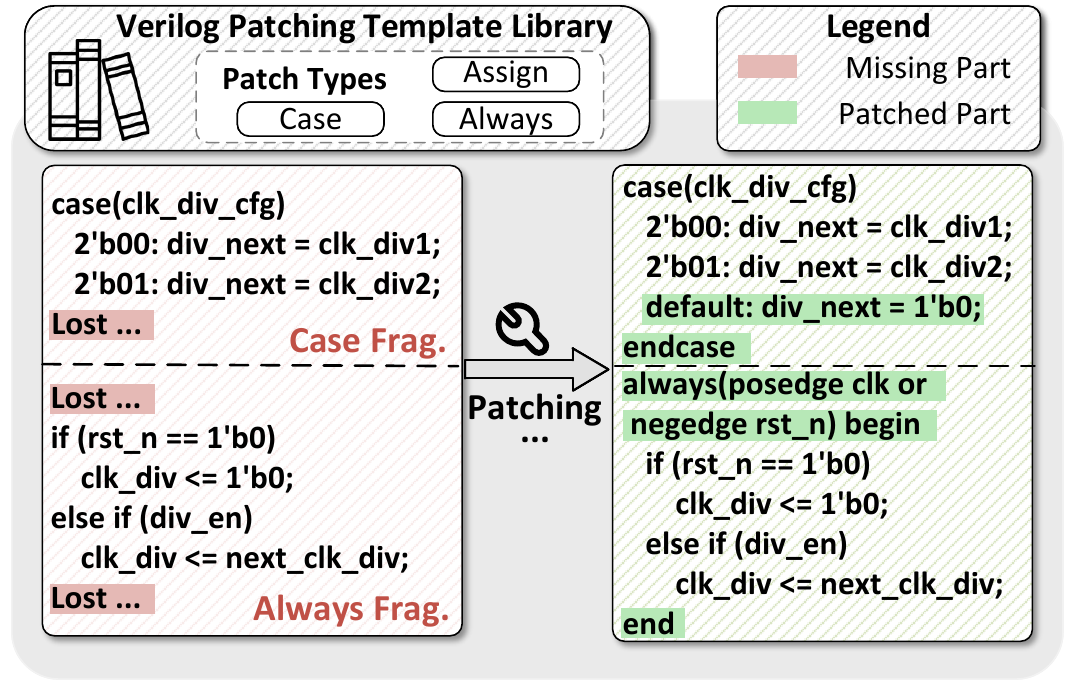} 
    \Description{Verilog Patching Template Library}
        \vspace{-10pt}
    \caption{\footnotesize
    \textbf{Verilog Patching Process.} The library reconstructs incomplete code fragments into valid blocks based on their syntax types. In the example, it appends missing \texttt{endcase} and \texttt{default} keywords to terminated case statements and wraps isolated logic fragments within \texttt{always} blocks containing appropriate sensitivity lists.}
    \vspace{-10pt}
    \label{fig:template}
\end{figure}

The workflow is as follows. We take the key statements reported by the Signal Tracker, and the Verilog Patcher inspects the surrounding RTL to determine whether each fragment came from a case block, always block or module shell, then selects a matching template from the library. It fills in the missing structure to produce a patched version of the code, namely the Filtered DUT. As illustrated in Fig.~\ref{fig:template}, the left side shows an incomplete case and always fragments (in red), while the right side shows the patched version with restored \texttt{default}/\texttt{endcase} keywords and an enclosing \texttt{always} block with an appropriate sensitivity list. This Filtered DUT is then presented to the LLMs in the stimuli refinement stage.

\setlength{\textfloatsep}{2pt}
\begin{table*}[t]
\vspace{-5pt}
    \sffamily
    \footnotesize
    \belowrulesep=0pt
    \aboverulesep=0pt
    \centering
        \caption{Benchmark designs used for evaluation, covering diverse hardware modules across multiple on-chip interfaces, including APB-based watchdog and power-control units, an AHB CORDIC accelerator, Q-Channel and P-Channel low-power controllers, and an AXI-based interface remap block.}
        \vspace{-10pt}
    \resizebox{1.0\linewidth}{!}{%
    \begin{tabular}{c|c|l|c|c}
    % {p{0.7cm} p{3.4cm}<{\centering} p{6.9cm}<{\centering}} 
    \toprule
     \textbf{Design Name} & \textbf{Protocol} & \textbf{Description} & \textbf{\makecell{Module\\ Counts}} & \textbf{Line Counts} \\

    \midrule
         \multirow{2}{*}{Watchdog}  & \multirow{2}{*}{APB}  & detects failures via a programmable counter, triggers an interrupt, and resets  & \multirow{2}{*}{3}   & \multirow{2}{*}{1500+} \\
         &&if ignored, with APB-configurable locked registers. & & \\\midrule
         Pwrctrl  & APB  & provides APB register control for SCP power, clock, and reset signal management.   & 9    & 2000+ \\\midrule
         Cordic   & AHB  & provides hardware acceleration for trigonometric and square root calculations.   & 3    & 1100+ \\\midrule
         \multirow{2}{*}{IdleControl}   & \multirow{2}{*}{Q-Channel}  & manages AXI and DMA interfaces, entering idle/stop states and accepting or  & \multirow{2}{*}{3}  & \multirow{2}{*}{800+}  \\
         & & rejecting low-power requests based on interface activity. & & \\\midrule
         LPctrl  & P-Channel  & enables safe CDC and low-power data exchange by using a low-power channel.    & 8    & 1500+ \\\midrule
         \multirow{2}{*}{Busremap}  & \multirow{2}{*}{AXI}  & converts master-slave signals, providing timing isolation, synchronisation, and  & \multirow{2}{*}{10}   & \multirow{2}{*}{3000+} \\
         & & transaction tracing for secure, efficient data exchange. & & \\   
         \bottomrule
    \end{tabular}}
    \vspace{-10pt}
    \label{tab:ben}
\end{table*}

\subsection{LLM-Guided Sequence Execution}
\label{subsc:SG}

In a UVM testbench, the sequence is the execution vehicle of stimuli: it decides which transactions are applied to the DUT and in what order and timing. In our framework, the LLMs reason on the Filtered DUT and uncovered points, but all generated sequences are instantiated and run on the original subsystem-level UVM testbench, so coverage is always measured on the original DUT.

To avoid relying on a single model’s judgment, we employ three different LLMs for coverage analysis and sequence generation, which helps reduce bias and explore a broader stimulus space. Given the Filtered DUT and uncovered items, the LLMs propose candidate sequences that are instantiated on the original DUT and simulated. During this loop, the models are also asked to flag coverage points that appear intrinsically unreachable; these are collected as waiver candidates and can be marked as waived in the final coverage report. If a candidate sequence causes compilation or simulation errors, the error messages are fed back to the LLMs to repair or discard the sequence and try again, closing the feedback loop of our framework.

% To ensure the correctness of waive decisions, we combine LLM-based semantic reasoning with formal verification, enabling cross-validation. The LLM provides high-level explanations for why a point may be unreachable, while the formal engine supplies provability results or counterexample traces. Both produce a waiving report that lists the underlying causes—such as signal constraints, infeasible paths, or restricted state transitions. These reports are then reviewed by verification experts, who make the final waive decision.

%\input{contents/Method1}
%\input{contents/Method2}
\vspace{-3pt}
\section{Evaluation}
\vspace{-3pt}
\label{sc:Evaluation}

% In this section, we present the experimental setup, evaluation metrics, and results.

\parlabel{Setup.}
We evaluate \textbf{\name}\ using LLM agents deployed through the ChatGPT API, with GPT-4.1 as the default model and Claude 4.5/Gemini 2.5 pro as comparative baselines. All generated testbenches are compiled and simulated using Synopsys VCS.

\vspace{-1.5pt}
\subsection{Evaluation Metrics}
\vspace{-3pt}

\paragraph{Success Rate of Generation (SRG)} This metric measures how often a generated UVM testbench is both syntactically valid and functionally correct. For each design, we let the LLMs generate $N_{\text{total}}$ testbenches. A generation is counted as successful if (i) the testbench completes the full VCS compilation and simulation flow without errors, and (ii) under the same regression stimuli, all checkers pass and the observed outputs match those of a reference constructed by experienced verification engineers. The success rate is
\vspace{-5pt}
\begin{equation}
\small
\label{eq:SRg}
\vspace{-7pt}
\mathbf{SRG} = \frac{N_{\text{correct}}}{N_{\text{total}}} \times 100\%.
\end{equation}

\paragraph{Coverage.}
After a testbench successfully runs on VCS, we report code coverage and functional coverage collected by the simulator.

\textbf{Code coverage} includes:
\vspace{-1pt}
\begin{itemize}
  \item \textbf{Score}: aggregate code coverage score for a design.
  \item \textbf{Line}: each executable line is hit at least once.
  \item \textbf{Branch}: both outcomes of each branch are taken.
  \item \textbf{Condition}: each Boolean expression is exercised.
  \item \textbf{Toggle}: each signal bit switches between 0 and 1.
\end{itemize}

\textbf{Functional coverage} is measured by user-defined covergroups and coverpoints that track whether specified scenarios, transactions and state combinations have been exercised. \footnote{
We focus on code coverage in our comparisons because it is defined directly on the RTL structure and is therefore comparable across designs and methods. Functional coverage instead depends on engineer-defined covergroups and reflects project-specific intent; in practice, it is only examined after code coverage exceeds about 90\%, and even then remains subjective. To keep our evaluation objective, all comparative and ablation studies use code coverage as the primary metric, and unless otherwise stated \textbf{coverage} refers to code coverage; we still report functional coverage for completeness in Table~\ref{tab:design_coverage}, using covergroups written by experienced engineers.}

% \paragraph{Completion Time.}
% We measure the time required to complete the full verification workflow—from initial generation to the final, simulation-validated testbench.

\begin{table}[t]
\sffamily
\footnotesize
\centering
\caption{Code coverage and functional coverage for each design.}
\label{tab:design_coverage}
\vspace{-10pt}
\begin{tabular}{ccc}
\toprule
\textbf{Design} & \textbf{Code Coverage (\%)} & \textbf{Functional Coverage (\%)} \\
\midrule
Watchdog    & 98.84 & 100    \\
Pwrctrl     & 93.66 & 90.64  \\
Cordic      & 100   & 100    \\
IdleControl & 94.90 & 96.12  \\
LPctrl      & 90.83 & 89.33  \\
Busremap    & 95.66 & 98.27  \\
\bottomrule
\end{tabular}
\end{table}
\vspace{-5pt}

\begin{figure}[b]    
\centering    
\vspace{-5pt}
\includegraphics[height=0.15\textheight,keepaspectratio]{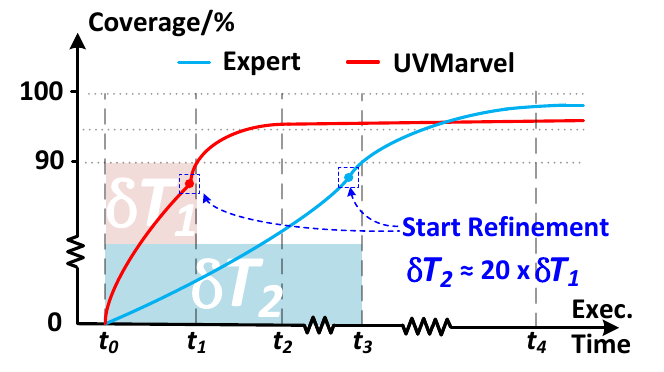}    
\Description{exe time}    
\vspace{-14pt}    
\caption{
\textbf{\footnotesize End-to-end verification time: \textbf{\name}\ vs. experts (across all benchmarks).} We exclude the test planning/IR authoring phase ($0$-$t_0$) from time evaluation to ensure precision—although this slightly lowers the acceleration ratio- to aligns UVMarvel’s automated verification objectives with manually defined coverage goals, guaranteeing fair and objective experimental results.
UVMarvel achieves 90\% coverage nearly twenty times faster than manual verification and converges to an industrial-grade final coverage level comparable to the expert flow. ($t_0$: end of IR/test planning, $t_1$: 90\% coverage by \textbf{\name}, $t_2$: coverage closure by \textbf{\name}, $t_3$: 90\% coverage by experts, $t_4$: coverage closure by experts, $\delta T_1$: time for \textbf{\name} to reach 90\% coverage, $\delta T_2$: time for experts to reach 90\% coverage)
}
\label{fig:runtime}
\end{figure}

\subsection{Benchmark}
Unlike IP-level benchmarks such as RTLLM \cite{lu2023rtllm}, Verilog-Eval \cite{liu2023verilogeval} and UVM² \cite{ye2025concept}, operating on isolated RTL blocks, our evaluation is based on industrial subsystem-level designs, reflecting genuine verification challenges. As summarised in Table~\ref{tab:ben}, the benchmarks span heterogeneous IPs connected through APB, AHB, AXI, P-Channel, and Q-Channel interfaces.
% , each requiring coordinated protocol handshakes, hierarchical register programming. 

\vspace{-5pt}
\subsection{Overall Framework Effectiveness}
\label{subsec:overall_effect}
\vspace{-3pt}

\textbf{Execution time.} \name\ covers both UVM testbench construction and stimuli refinement. Starting from a human-provided IR and running until code coverage reaches 90\%, our framework takes 4.5 hours on the benchmark subsystem, which is 20.17$\times$ faster than the manual verification flow, as shown in Fig.~\ref{fig:runtime}. In other words, subsystem-level verification tasks that previously required several working days of human-driven UVM development and regression are now completed within an hour-scale automated run.

\textbf{SRG of testbench generation.} \name\ achieves an overall SRG of 93.33\% across all subsystem components: only AXI-based subsystems fail, mainly due to incorrect coordination between AXI channels (e.g., inconsistent AW/W/B handshakes). In contrast, existing testbench-generation work (e.g., MEIC\cite{xu2024meic} and UVM$^2$\cite{ye2025concept}) achieves 0\% SRG at it, because these methods are limited to IP-level RTL and cannot handle register configuration, bus transactions, and other issues that arise only in subsystem-level RTL.

\begin{figure*}[t]
    \centering

    % Fig.6（最上面横着）
    \begin{minipage}[t]{1\textwidth}
        \centering
        \vspace{-15pt}
        \includegraphics[trim=0mm 0mm 0mm 0mm, clip, width=\linewidth]{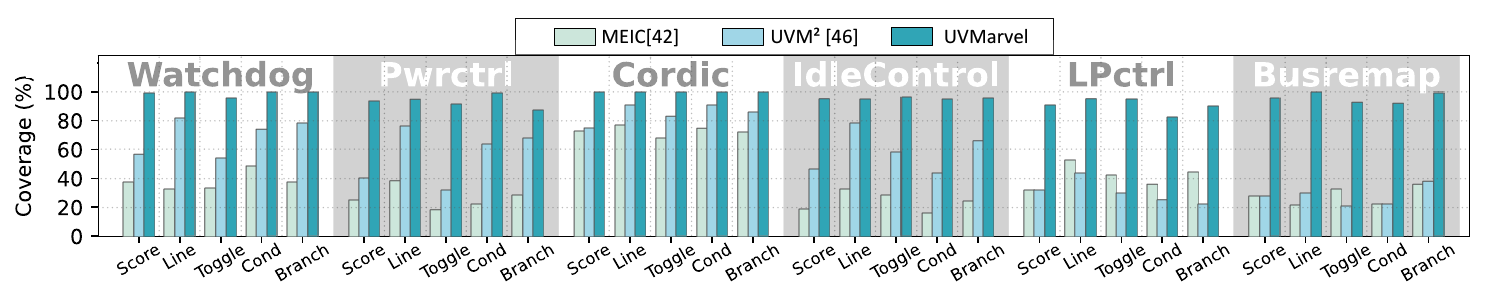}
        \vspace{-20pt}
        \caption{Code coverage of MEIC \cite{xu2024meic}, UVM² \cite{ye2025concept}, and UVMarvel across six benchmark tests, evaluated using five coverage components.}
        \label{fig:ex1}
    \end{minipage}

    \vspace{-1pt} % 图6和下面一排之间的垂直间距

    % Fig.7 和 Fig.8 并排
    \begin{minipage}[t]{0.49\textwidth}
        \centering
        \includegraphics[trim=0mm 0mm 0mm 0mm, clip, width=\linewidth]{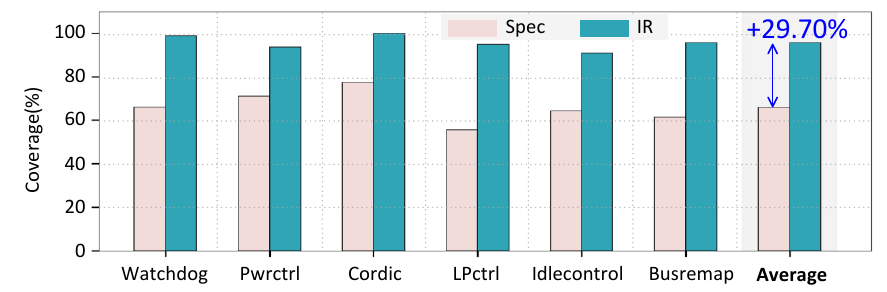}
        \vspace{-20pt}
        \caption{Code coverage comparison between IR and SPEC inputs across six benchmark tests, showing higher coverage achieved by IR inputs.}
        \label{fig:ex2}
    \end{minipage}
    \hfill
    \begin{minipage}[t]{0.49\textwidth}
        \centering
        \includegraphics[trim=0mm 0mm 0mm 0mm, clip, width=\linewidth]{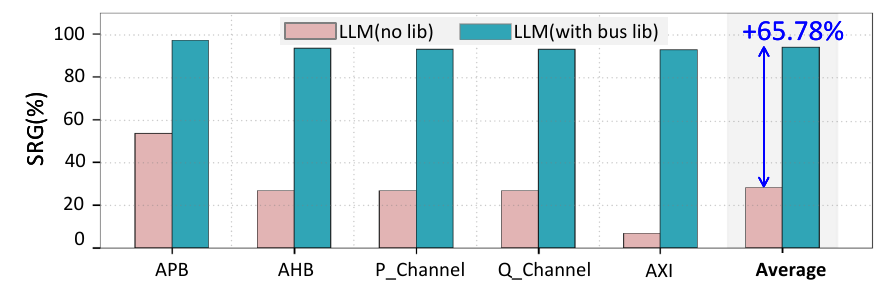}
        \vspace{-20pt}
        \caption{Comparison of SRG performance with and without the Bus Protocol Library, validated over five different bus protocols.}
        \vspace{-25pt}
        \label{fig:ex3}
    \end{minipage}

\end{figure*}

\textbf{Coverage.} \name\ reaches  95.65\% average code coverage. We compare against two automated stimuli-generation baselines: MEIC\cite{xu2024meic}, relying on random stimuli, and UVM$^2$\cite{ye2025concept}, activating LLM capabilities through structured prompts and templates. As both lack bus-protocol knowledge, we provide them with a pre-constructed subsystem-level UVM testbench and Bus Protocol Library used by \name\, so that the comparison isolates the stimuli generation capability. As shown in Fig.~\ref{fig:ex1}, \name\ achieves higher code coverage on all metrics in our benchmark.
% with score coverage on average 57.15\% above MEIC\cite{xu2024meic} and 48.23\% above UVM$^2$\cite{ye2025concept}. These suggest that existing stimuli refinement, while effective at the IP level, has not yet closed coverage at the subsystem level.

\vspace{-3.5pt}
\subsection{Impact of IR on Stage \circled{a}}
\label{subsec:ex_IR}
\vspace{-3pt}

We evaluate the impact of the IR by comparing two settings: giving the LLMs the original specification and giving them the IR. As shown in Fig.~\ref{fig:ex2}, using the IR improves coverage by nearly 30\% on average, indicating that the IR is more than a simple paraphrase of the specification. On closer inspection, we observe that in the spec-only setting, the LLMs often instantiate incomplete environments, for example, omitting bus monitors or failing to connect register models to the corresponding bus agents, so that large parts of the DUT are never driven. By presenting design intent in a predictable, composable format, the IR helps the model infer handshake directions, transaction ordering and interface semantics, leading to more complete and effective subsystem-level UVM testbenches.
% \begin{figure}[t]
%     \centering
%     \includegraphics[trim=00mm 00mm 00mm 00mm, clip, width=1\linewidth]{graphics/png1.1/ex3_2.png} 
%     \vspace{-20pt}
%     \caption{ lib or not
%     }
%     \vspace{-5pt}
%     \label{fig:ex3} % 便于后续引用
% \end{figure}

\vspace{-3pt}
\subsection{Impact of Bus Protocol Library on Stage \circled{a}}
\label{subsc:ex_bus}
\vspace{-3pt}
We further evaluate the role of the Bus Protocol Library by comparing two settings: (1) prompting the LLMs without the library, and (2) supplying the library as guidance. In both cases, all non-bus UVM components are fixed and provided as context; the LLMs are only asked to generate the bus-side components. 

As shown in Fig. \ref{fig:ex3}, providing the Bus Protocol Library improves the SRG by 65.78\%, demonstrating that the LLMs struggle to infer protocol behaviours directly from specifications. The challenge is particularly acute for AXI, where handshake patterns and decoupled channels introduce nontrivial ordering, response, and backpressure constraints. Without guidance, the success rate for AXI falls below 7\%; with the protocol library, it exceeds 90\%. 
% These results show that the Bus Protocol Library effectively equips the LLMs with the ability to construct correct bus-side UVM components.

% These results reinforce that the Bus Protocol Library is not merely a code repository. Instead, it functions as a semantic compression of protocol behaviour, distilling hundreds of pages of protocol manuals into a compact, executable template that the LLMs can reliably adapt. By externalising protocol semantics, the library enables consistent, correct construction of subsystem-level bus components.

\vspace{-3pt}
\subsection{Impact of Signal Tracker \& Verilog Patcher on Stage \circled{b}}
\label{subsc:ex_stimuli}
\vspace{-3pt}

\begin{figure}[t]
    \centering
    \vspace{-15pt}
    \includegraphics[trim=00mm 00mm 00mm 00mm, clip, width=1\linewidth]{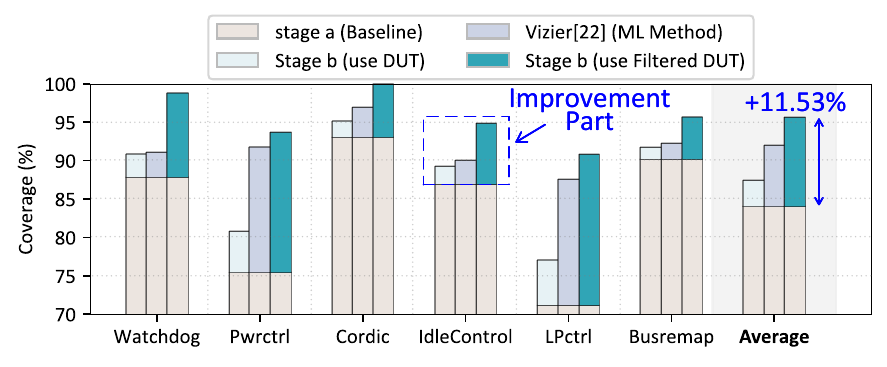} 
    \vspace{-25pt}
    \caption{ Coverage improvement achieved by the three methods across six benchmark tests, with the filtered DUT getting the highest performance.
    \hspace{5pt}
    \vspace{-5pt}
    }
    \label{fig:ex4} % 便于后续引用
\end{figure}

We evaluate the effect of the Signal Tracker and Verilog Patcher by measuring score coverage improvement during refinement. Prior work has shown that machine learning can assist in exploring design and parameter spaces~\cite{chen2024deep,wang2025automated,lu2025deepverifier,vasudevan2021learning,tweehuysen2023stimuli,huang2022test,ohana2023closing,dhodhi2021deep,samarah2006automated,abd2021speed,choi2021application,fajcik2017automation,bhargav2021enhancing,danciu2022coverage,gadde2024efficient,mondol2024rl}, so we also compare against a representative ML-based optimisation approach. Concretely, we adopt Google’s open-source Vizier framework~\cite{huang2022test} as an ML-based refinement method.

% Using the coverage of Stage~\circled{a} as the baseline, we compare three refinement strategies: feeding the LLMs the full DUT, refining stimuli with Vizier, and feeding the LLMs the Filtered DUT. As shown in Fig.~\ref{fig:ex4}, the Filtered DUT achieves the largest average gain (11.53\%), while Vizier~\cite{huang2022test} and the raw DUT yield only 6.66\% and 3.35\%, respectively. These results lead to two observations. (1) When applied directly to the full subsystem RTL, the LLMs make limited progress in improving coverage; even though Vizier~\cite{huang2022test} requires many optimisation iterations and extra compute, it copes better with the large search space than unguided LLM refinement. (2) Once we provide compact and dependency-preserving context through the Filtered DUT, the LLMs surpass both the raw-DUT and ML baselines in coverage gain, suggesting that LLMs have strong potential for verification when combined with appropriate structural guidance rather than relying on prompt engineering alone.

Using the coverage of Stage~\circled{a} as the baseline, we compare three refinement strategies: feeding the LLMs the full DUT, refining stimuli with Vizier, and feeding the LLMs the Filtered DUT. As shown in Fig.~\ref{fig:ex4}, the Filtered DUT achieves the largest average gain (11.53\%), while Vizier~\cite{huang2022test} and the raw DUT yield only 6.66\% and 3.35\%, respectively. These results lead to two observations. (1) When applied directly to the full subsystem-level RTL, the LLMs make limited progress in improving coverage; even though Vizier~\cite{huang2022test} requires many optimisation iterations and computation, it copes better with the large search space than unguided LLM refinement. (2) Once we provide compact and dependency-preserving context through the Filtered DUT, the LLMs surpass both the raw-DUT and ML baselines in coverage gain, suggesting that the LLMs have potential for verification when combined with appropriate structural guidance.
% rather than relying on prompt engineering alone.

\begin{table}[t]
\sffamily
\footnotesize
\centering
\vspace{-10pt}
\caption{Code coverage achieved by different LLMs.}
\label{tab:pipeline_rationality}
\vspace{-10pt}
\begin{tabular}{cccc}
\toprule
\textbf{Design Name} & \textbf{GPT4.1} & \textbf{Claude4.5} & \textbf{Gemini2.5pro} \\
\midrule
Watchdog    & 98.70 & 98.70 & 98.84 \\
Pwrctrl     & 93.40 & 93.66 & 93.66 \\
Cordic      & 100.00 & 100.00 & 100.00 \\
IdleControl & 94.90 & 94.80 & 94.60 \\
LPctrl      & 90.70 & 90.70 & 90.83 \\
Busremap    & 95.40 & 95.66 & 95.60 \\
\bottomrule
\end{tabular}
\end{table}

\vspace{-5pt}
\subsection{Code coverage achieved by different LLMs}

\vspace{-1.5pt}
To reduce the risk of relying on a single model that might miss corner cases, Stage~\circled{b} employs three different LLMs when generating new sequences. As shown in Table~\ref{tab:pipeline_rationality}, the three models achieve very similar coverage under the same IR, protocol scaffolding and Filtered DUT. This convergence across models suggests that UVMarvel is robust to the choice of the LLMs; its coverage improvement remains stable even when the underlying model changes.

\vspace{-5pt}
\section{Conclusion}
\label{sc:Conclusion}
 We have presented \name\ , the first UVM-based verification framework for subsystem-level RTL. By combining the IR with the Bus Protocol Library, UVMarvel automatically constructs subsystem-level UVM testbenches, and by using the Signal Tracker together with the Verilog Patching Library, it further boosts score coverage to 95.65\% with a 4.5-hour automated execution.
\section{Acknowledgement}
\label{sc:acknowledgement}
We appreciate the reviewers for their helpful feedback.
This work is supported by the National Key Research and Development Program (Grant No.2024YFB4405600), the Basic Research Program of Jiangsu (Grants No. BK20243042), and the Fundamental Research Funds for the Central Universities (No. 2242025K20013).

% \clearpage
% \bibliographystyle{IEEEtran}
% \bibliography{ref}

% \end{document}

% \clearpage

\bibliographystyle{ACM-Reference-Format}
\bibliography{sample-base}

\balance
\end{document}